# Distilling the Essence of an Evolutionary Process and Implications for a Formal Description of Culture

LIANE GABORA
Department of Psychology, University of California, Berkeley, USA
and
Center Leo Apostel for Interdisciplinary Studies, Vrije Universiteit Brussel, Brussels, Belgium

and

DIEDERIK AERTS
Center Leo Apostel for Interdisciplinary Studies and Department of Mathematics
Vrije Universiteit Brussel, Brussels, Belgium

It has been proposed that, since the origin of life and the ensuing evolution of biological species, a second evolutionary process has appeared on our planet. It is the evolution of culture—*e.g*. ideas, beliefs, and artifacts—and the creative minds that invent them, adapt them to new situations, and play with them for artistic expression and fun. But does culture evolve in the same genuine sense as biological life? And if so, does it evolve through natural selection, or by some other means? Why does no other species remotely approach the degree of cultural complexity of humans? These are questions that must be addressed because they lie at the foundation of who we are and what makes our lives meaningful.

Although much research has been done on how selective pressures operating at the biological level affect cognition and culture, little research has focused on culture as an evolutionary process in its own right. Nonetheless, culture *does* appear to evolve. Like biological forms, cultural forms—ideas, attitudes, artifacts, mannerisms, *etc.*—incrementally adapt to the constraints and affordances of their environment through descent with modification. Agricultural techniques become more efficient, computers get faster, scientific theories predict and account for more observations, new designs are often artistic spin-offs of those that preceded them. And in some respects culture appears to be Darwinian, that is, a process of differential replication and selection amongst randomly generated variants. For example, different brands of peanut butter may be said to compete to be 'selected' by consumers. This suggests that knowledge of biological evolution can be put to use to gain insight into cultural patterns. However, the attempt to straightforwardly apply Darwinian theory to culture has not been overwhelmingly fruitful. It certainly hasn't provided the kind of unifying framework for the social sciences that Darwin's idea of natural selection provided for the biological sciences. This is largely because the underlying substrate of the process—human beings—are notoriously complex and unpredictable! For example, natural selection cannot tell us much about how someone came up with the idea for turning peanuts into a spreadable substance in the first place!

The difficulty applying evolutionary theory as it has been developed in biology to culture arises largely because of the highly nonrandom manner in which the mind—the hub of cultural change—generates and assimilates novelty. To understand how, when, and why the human mind became capable of supporting culture, and what may have previously held it back, we need to know something about how we attained the creative powers we now possess, and how creative processes actually work, in groups as well as individuals. To invent in the strategic, intuitive manner characteristic of the human mind requires a cognitive architecture that supports the capacity to spontaneously adapt concepts to new circumstances and merge them together to conceptualize new situations. Thus we find that at the heart of the puzzle of how culture evolves lies the *problem of concepts*, not so much just how we use them to identify and classify objects in the world, but their contextuality and compositionality, and the creative processes thereby enabled.

We will see that the change-of-state a mind undergoes as it develops an idea is not a natural selection process, and indeed it may be that culture evolves, but only in small part through Darwinian mechanisms. We suggest that its basic mode of evolving turns out to be a more general process referred to as *context-driven actualization of potential*. Thus the story of how ideas are born and bred in one mind after another leads us to another story, that of what it means to evolve, and how an evolutionary process could work. Finally, this paper will touch on how an evolutionary perspective on culture can shed light on questions of a philosophical or spiritual nature that have been with us since the first fledgling creative insights glimmered in our ancestors' brains.

## 1 Do Evolutionary Models Capture the Dynamics of Culture?

Let us consider how well attempts to formally or informally describe culture as an evolutionary process do at capturing the cultural dynamic.

### *1.1 Memes*

Perhaps the most well known attempt to apply Darwinism to culture is the meme approach (Aunger 2000; Blackmore 1999, 2000; Dawkins 1976). It simplifies things by restricting what counts as 'culturally transmitted' to things that are passed from one person to another relatively intact, such as eye-catching fashions, or belief in God. This approach quickly runs into problems. First because ideas and stories are not simply stored, outputted, and copied by others as discreet chunks, complete unto themselves. They are dynamically influenced by the context in which they appear, and we process and re-process them in ways that reflect our unique experiences and unique style of weaving them into an internal model of the world, or *worldview*. Furthermore, the meme perspective leads us to view ourselves as 'meme hosts', passive imitators and transmitters of memes. Although some authors have capitalized on the shock value of the ensuing dismal view of the human condition, clearly we are not merely passive hosts but active evolvers of culture.

### *1.2 Mathematical Approaches*

Others have drawn from mathematical models of population genetics and epidemiology to model the spread of ideas (Cavalli-Sforza & Feldman, 1981; Schuster & Sigmund, 1983; Boyd & Richerson, 1985). They examine the conditions under which mutated units of culture pass vertically via family, or horizontally through a community by imitation within an age cohort, and proliferate. The limitations of this approach are expressed succinctly by Kauffman (1999):

> True, but impoverished. Why impoverished? Because the concept of meme, and its descent with modification is taken as a, or perhaps 'the' central conceptual contribution to the evolution of human culture. But the conceptual framework is so limited as to be nearly trivial. Like NeoDarwinism, it suffers from the inability to account for the source of new forms, new memes. Further, mere descent with modification is a vastly oversimplified image.
>
> Consider the new concepts, artifacts, legal systems, modes of governance, modes of coevolving organizations at different levels that have come into existence in the past three million years. Our understanding of these and other aspects of culture transforms every day. Take, for instance, the Wright brother's airplane. It is a recombination of four technological facts: an airfoil, a light gas engine, bicycle wheels, and a propeller. The more diversity that exists in a technological community, the more diversity of novel combinations of existing elements are present that might later prove useful in some context. Thus, 200,000 years ago, the diversity of the economic web of goods and services was severely limited. Today it is vast. 200,000 years ago, finding a technological novelty with the stone and bone implements available was hard. Today, with millions of artifacts already in existence, the generation of novel ones is easy.
>
> In short, memes do not just descend with modification. A rich web of conceptual interactions is at work as humans happen upon, design, and implement a combinatorially exploding diversity of new goods and services. This WEB structure of technological and cultural evolution is far richer, and far closer to the truth, than mere meme descent with modification. Indeed, this broader view helps us begin to understand how and why memes recombine and diversify. It is a more generative picture, undoubtedly still inadequate, but far better than a naïve copying of neoDarwinism.

### *1.3 Computer Models*

To what extent we can computationally abstract the underlying skeleton of the cultural process and actually evolve something with it? If culture, like biology, is a form of evolution, it should be possible to develop a minimal model of it analogous to the *genetic algorithm*, a biologically inspired search tool that evolves solutions to complex problems through a reiterated process of randomly mutating information patterns and selectively replicating those that come closest to a solution (Holland 1975).

*Meme and Variations* (or MAV for short) is to our knowledge the first computer model of the process by which culture evolves in a society of interacting individuals. It is discussed only briefly here since it is presented in detail elsewhere (Gabora 1995). MAV consists of an artificial society of neural network-based agents that don't have genomes, and neither die nor have offspring, but that can invent, assess, imitate, and implement ideas, and thereby gradually increase the fitness of their actions. Agents have an unsophisticated but functional capacity to *mentally simulate* or assess the relative fitness of an action before actually implementing it (and this capacity can be turned off). They are also able to invent strategically and intuitively, as opposed to randomly, building up 'hunches' based on trends that worked in the past (and this too can be turned off). This was possible because of the integrated structure of the neural network. All the agents' concepts are connected, if indirectly, to one another, and thus each can influence, if only weakly, each other. The architecture of MAV is also such that it implements a cultural version of *epistasis*. In biological epistasis, the fitness conferred by one gene depends on which allele is present at another gene. In MAV, the fitness conferred by the locus determining the movement of one limb depends on what the other limbs are doing.

Initially all agents are immobile. Every iteration, each agent has the opportunity to acquire a new idea for some action, either through 1) *innovation*, by strategically modifying a previously learned idea, or 2*) imitation*, by copying an action performed by a neighbor. Quickly some agent invents an action that has a higher fitness than doing nothing, and this action gets imitated by others. As ideas continue to be invented, assessed, implemented as actions, and spread through imitation, the diversity of actions increases. Diversity then decreases as the society evolves toward implementing only those actions that are most fit.

MAV exhibits many phenomena observed in biology, such as *drift*—changes in the relative frequencies of different alleles (forms of a gene) as a statistical byproduct of randomly sampling from a finite population. Second, as in biology we find that epistasis increases the amount of time it takes to evolve. Third, although in the absence of variation-generating operations culture does not evolve, increasing innovation much beyond the minimum necessary causes average fitness to decrease, just as in biology.

MAV also addresses the evolutionary consequences of phenomena unique to culture. Imitation, mental simulation, and strategic (as opposed to random) generation of variation all increase the rate at which fitter actions evolve. The higher the ratio of innovation to imitation, the greater the diversity, and the higher the fitness of the fittest action. Interestingly however, for the society as a whole, the optimal innovation-to-imitation ratio was approximately 2:1 (but diversity is compromised). For the agent with the fittest behavior, the less it imitated (*i.e.* the more effort reserved for innovation), the better. This suggests if you're the smartest one around, don't waste time copying what others are doing!

Thus it is possible to genuinely evolve information using a computer algorithm that mimics the mechanics of culture[1]. More recent computer models of cultural evolution (*e.g.* Spector & Luke, 1996a, b; Baldassarre, 2001) embed the cultural dynamic in a genetic algorithm. Thus agents not only exchange ideas but bear offspring and die. Although these models have unearthed interesting results concerning the interaction between biological and cultural evolution, we believe the first priority is to first learn what we can through computer simulations of culture alone before combining the two. After all, culture is not merely an extension of biology. Biology does not provide adequate explanatory power to account for the existence of widgets (just as physics cannot explain the existence of worms). Culture is spectacularly unlike anything else biological processes have given rise to. Indeed there is much left to do with such a culture-only modeling approach. Everyday experience suggests that human culture exhibits other phenomena observed in biological evolution that could be investigated with this kind of computer model, such as Founder Effect (stabilization in a closed-off social group) and altruism (being especially nice to those who are related to you). In fact one could argue that humans feel more altruistic toward their 'cultural kin' than their biological kin. (For example, who would you go out of your way for the most: someone who has the same eye color or blood type as you, or someone who shares your interests?)

---

[1] MAV will be elaborated such that agents have a more realistic method of generating novelty, and multiple drives that are satisfied to different degrees by different actions, and the fitness function for the evaluation of an idea emerges from the drive strengths.

### *1.4 Where Do We Stand?*

How well have we done at capturing what really happens in cultural evolution? At best, invention and imitation are modeled as single-step processes, in no way coming close to what really happens as a novel idea is churned through. There is a saying, 'you never step into the same stream twice', and it applies to streams of thought as well as streams of water. Units of culture are not retrieved whole and discreet from memory like apples from a box. Humans not only have the *ability* to blend and adapt ideas to new situations and see them in new perspectives, we are *compelled* to. And we are compelled to entice others to see things our way too, or to bat ideas around with one another, using each other as a mental scaffold. Moreover, just about anything is food for thought, and thus food for culturally-transmittable behavior. Some items in memory, such as a recipe for goulash, may be straightforwardly transmitted through imitation. Others, such as, say, an attitude of racial prejudice, appear to be culturally transmitted, but it is impossible to point to any particular phrase or gesture through which this transmission is mediated. Still others partake in the cultural dynamic in even subtler ways, as when a composer releases the painful experience of his daughter's death in a piece of music.

As an idea passes from one individual to another, it assimilates into the various minds it encounters, and these minds are altered to accommodate not only the idea but also what it may, perhaps only subtly, imply or suggest. An idea has a different impact on different individuals, depending on the beliefs and preconceptions already in place. Furthermore, individuals differ in the extent to which they process it, and thus the extent to which their worldview is affected by it and by its 'halo' of implications. They also differ in the extent to which their processing of the idea takes place alone or through interaction with others. There are individuals who are never directly exposed to the idea, but indirectly altered by it nevertheless, through exposure to others who *are* directly exposed. In short, the evolution of the ideas, stories, and artifacts that constitute culture is a subtle matter.

## 2 Background from Cognitive Science

In order to understand the mechanics of the creative thought processes through which cultural novelty is generated let us briefly examine the underlying cognitive architecture.

### *2.1 Conceptual Space and the Distributed Nature of Memory*

We begin by looking at how episodes or items of experience are stored in memory, and how we navigate through memory by way of abstract concepts. Episodes stored in the mind are *distributed* across a cell assembly that contains many locations, and likewise, each location participates in the storage of many items (Hinton *et al*., 1986; Palm, 1980). According to the doctrine of *neural re-entrance*, the same memory locations get used and reused again and again (Edelman, 1987). Each location is sensitive to a broad range of *subsymbolic microfeatures* (Smolensky, 1988), or values of them (*e.g.*, Churchland & Sejnowski, 1992). Thus location *A* may respond preferentially to lines oriented at say 45 degrees from the horizontal, neighboring location *B* to lines at a slightly different angle, say 46 degrees, and so forth. However, although *A* responds *maximally* to lines of 45 degrees, it responds to a lesser degree to lines of 46 degrees. This kind of organization is referred to as *coarse coding*. Thus when we speak of a distributed memory we speak of one where, for example, location *A* would participate in the storage of all those memories involving lines at close to 45 degrees, and each of these memories affect not just location *A* but a constellation of other locations. This kind of architecture is also said to be *content-addressable* because similar or related items activate, and get stored in, overlapping memory regions. It can be useful to think of a state of the mind as consisting of a specific combination of distinguishable features or *properties*, and of all possible states of mind as defining what can be called *conceptual space*.

This kind of distributed, content-addressable architecture enables culturally acquired information or past experiences to be recursively manipulated or *redescribed* in streams of thought, or indirectly colored by more recent events. The alterations they acquire as they are mulled over are highly nonrandom, reflecting not just the mind's analytic capacities but also its associative structure. And because the process is affected by the circumstances of the present moment, even simple recollection is a contextual *reconstructive* event.

### *2.2 Conceptual Integration*

Following the pioneering efforts of Piaget, Vygotsky and others, it has become clear that a worldview is not present from birth but develops naturally through experience in the world. The infant mind is predisposed to selectively attend biologically salient features, and respond accordingly. If it is hungry and sees its mother's breast, it suckles; if it feels something extremely hot or cold it pulls away, and so forth. In addition to innate predispositions to respond categorically to certain stimuli, it is widely thought that infants possess higher cognitive competencies (Gelman, 1993; Keil, 1995). These competencies may be due to core knowledge (Spelke, 2000), intuitive theories (Carey, 1985), or simply predispositions to direct attention to salient (particularly social) elements of a situation (Leslie, 2000). An infant is also capable of storing episodes as memories. Although episodes from infancy are rarely accessible later in life, they *do* get etched into memory, as evidenced by the capacity for reminding events, which is present by two month of age (Davis & Rovee-Collier, 1983; Rovee-Collier *et al*., 1999; Matzel *et al*., 1992).

Although the issue is controversial, it is widely accepted that between six and eight years of age, a child moves from implicit, domain-specific representations to explicit, more broadly applicable representations (Karmiloff-Smith, 1990, 1992). This is evidence of starting to have a sense of how the various aspects of life, society, and the world at large, fit together and relate to one another. Aided by social exchange, and mediated by artifacts, a framework for how things are and how things work falls into place, and it bears some likeness (and also some dis-similarities) to that of its predecessors, such as the worldviews of parents and other influential individuals. Some experiences are either so consistent, or so inconsistent with the worldview that they have little impact on it. Others mesh readily with existing ideas, or ring true intuitively, and percolate deep into the newly emerging worldview, renewing the child's understanding of a myriad other notions or events. The child is thereby encultured, becomes a unique cog in the culture-evolving machinery.

### *2.3 Focusing and Defocusing*

Once one has an integrated model of the world how does it get creatively put to use? Human thought processes vary along a continuum from rigorous and analytical to intuitive and associative (Ashby & Ell, 2002; Freud, 1949; James, 1890/1950, Johnson-Laird, 1983; Kris 1952; Neisser, 1963; Rips, 2001; Sloman, 1996), and it has been experimentally demonstrated that particularly creative individuals excel at both (Barron, 1963; Eysenck, 1995; Feist, 1999; Fodor, 1995; Richards *et al*., 1988; Russ, 1993). Accordingly, it has been proposed that creativity involves the ability to subconsciously focus or defocus attention, thereby varying the size of the memory region impacted by and retrieved from in response to a situation (Finke *et al*., 1992; Gabora, 2002; Martindale, 1995). This capacity, referred to as *contextual focus*, enables one to alternate between *analytical thought*, where the impacted region is small enough to zero in on only the most relevant or defining aspects of a situation, and *associative thought*, where it is large enough that seemingly less relevant aspects come into play. When attempts to solve a problematic situation analytically are unsuccessful, attention becomes defocused, and one takes more aspects into account. Thought becomes more associative, which may throw a new perspective on the situation. Maintaining this new perspective while resuming a state of focused attention—where mental effort is reserved for the sort of complex operations characteristic of analytic thought—may lead to a solution.

## 3 Evolution of the Culture-evolving Mind

Having examined the fluid nature of our novelty generating abilities, we are ready to consider: how did these abilities come about? In this section we speculate about how the creative cognitive structure described in the previous section could have come about, drawing on evidence from archeology and anthropology.

### *3.1 What Sparked the Origin of Culture?*

The origin of task-specific tools, organized hunting, fire use, and migration out of Africa 1.7 million years ago are suggestive of a significant cognitive transition at this time. It is proposed that this transition occurred due to neurophysiological changes enabling the receptive fields where memories are storied and retrieved from to become more distributed, facilitating reminding events and concept formation. This would have paved the way for onset of the capacity for memories and concepts to become integrated through the formation of a dynamical network of concepts to yield a self-modifying worldview. It is further proposed that the process of conceptual integration comes about through a process referred to as *conceptual closure* (Gabora, 2002, submitted). The notion of a closure space comes from a branch of topology known as graph theory. It deals with how *points* can be connected by *edges*, and the basic idea can be explained easily as follows. Imagine you have a jar full of buttons, which you spill on the floor. You tie two randomly chosen buttons with a thread, and repeat this again and again. Occasionally you lift a button and see how many connected buttons get lifted, and you find that clusters start to emerge. When the ratio of strings to buttons reaches about 0.5, you arrive at what is called a *percolation threshold*, where clusters of connected buttons join to form a giant cluster containing most of the buttons (Erdos & Renyi, 1959, 1960; Kauffman, 1993). Thus closure in the mathematical sense does not mean that nothing can get in or out. It means that there exists a path for getting from any one point to any other in the set by means of connected points.

Now we apply the concept of closure to cognition. Memories are described as points (buttons), associative paths between them as edges (strings), and concepts as clusters of connected points. Learning and reminding increase the density of associative paths, and the probability of concept formation. Concepts facilitate streams of thought, which forge connections between more distantly related clusters. The ratio of associative paths to concepts increases until it becomes almost inevitable that one giant cluster emerges and the points form a connected closure space. There now exists a potential associative pathway from any one memory or concept to any other. Because the memory is integrated, it has the capacity to reason about one thing in terms of another, adapt ideas to new circumstances, or frame new experiences in terms of previous ones, and combine information from different domains as in a joke. It should be stressed that it is not the *presence of* but the *capacity for* an integrated worldview that the human species came to possess. An infant may be born predisposed toward conceptual integration, but the process must begin anew in each young mind.

### *3.2 The Earliest Modern Minds and the 'Cultural Revolution'*

So we have a worldview in which different domains can be associated at an abstract level. But how did we come to have the ability to discern and analyze what are the relevant aspects of these associations so as to make efficient use of them?

A second cultural transition took place approximately 50,000 ka, during the Upper Paleolithic. We see at this time a more strategic style of hunting involving specific animals at specific sites, colonization of Australia, replacement of Levallois tool technology by blade cores in the Near East, elaborate burial sites indicative of ritualized religion, and the first appearance of art, jewelery, and decoration of tools and pottery in Europe. There is also evidence of modern language, and a restructuring of social relations. Moreover, cultural change becomes cumulative, one change building on another, what has come to be called the Ratchet Effect (Tomasello 1993).

It has been suggested that what was necessary to bootstrap culture was the capacity for a theory of mind (ToM), which refers to the capacity to reason about the mental states of others. However, ToM is not the golden egg; to be able to strategically invent, refine, and communicate, much more is involved. It is proposed that this transition resulted from fine-tuning of the mechanisms underlying *contextual focus*, which as we saw earlier is the capacity to subconsciously focus or defocus attention in response to the situation, thereby varying the size and diversity of the memory region impacted by and retrieved from (Gabora, 2003). This enabled humans to spontaneously shift between analytic and associative modes of thought.

Once we acquired the capacity for contextual focus, when attempts to solve a problem analytically were unsuccessful we could defocus attention, enter a more associative form of thought, and see it in a new light. Resuming a state of focused attention, but now viewing the problem in a new way might lead to a solution. If not, the focus/defocus process could be repeated. Thus it became possible to generate new approaches to the myriad obstacles and dilemmas large and small that arise in everyday life. The onset of contextual focus could have given rise to the capacity for conceptual closure at multiple hierarchical levels, enabling each potential element of culture to be viewed from different perspectives within a continually changing integrated conceptual framework. It is proposed thus to have played in the origin of art, science, religion, and possibly modern language.

## 4 Rethinking Evolution

Now that we have an integrated worldview capable of manifesting and refining contextually relevant actions and artifacts, let us return to the theoretical issue of how culture evolves. Is there replicator in cultural evolution, and if so what is it? Do worldviews evolve like biological organisms through a natural selection process, or by some other means? What makes something count as an evolutionary process in the first place?

### *4.1 The Cultural Replicator: Minds Not Memes*

It is often assumed that the basic units of cultural evolution are artifacts like tools, fashions, and so forth, or the mental representations or ideas that give rise to these concrete cultural forms. Moreover, it is suggested that artifacts or ideas constitute 'replicators', cultural entities that replicate themselves in the same sense as living organisms do.

The concept of replicators was thought through deeply by von Neumann. He postulated that a genuine self-replicating system consists of coded information that can and does get used in two distinct ways (von Neumann 1966). One way is as merely a description of itself, or *self-description*, that is *passively copied* to the next replicant. In this case, the code is said to be used as *uninterpreted information.* The other way is as a set of instructions for how to put together a copy of itself; that is, as *self-assembly instructions* that are *actively deciphered* to build the new replicant. In this case, the code is said to be used as *interpreted information.* To put it more loosely, the interpreting process can be thought of as 'now we make a body', and the uninterpreted use of the code as 'now we make something that can *itself* make a body'. Since biology is the field that inspired this distinction, naturally it applies here. The DNA self-assembly code is copied—without interpretation—to produce new strands of identical DNA during the process of meiosis. In successful gametes, these strands of DNA are decoded—that is, interpreted—to synthesize the proteins necessary to construct a body during the process of development.

Neither an artifact nor an idea is a replicator in the strict sense identified by von Neumann because it does not consist of self-assembly instructions. It may *retain* structure as it passes from one individual to another, but does not *replicate* it. Its transmission is more akin the transmission of a radio signal and its reception by one or more radios; neither an idea nor a radio signal self-replicates in the biological sense, copying and interpreting an explicit self-assembly code. Thus it has been argued that the cultural replicator is not an idea but a conceptually closed web of them that together form a mind, or from an 'inside' point of view, a worldview (Gabora, 2004). A worldview replicates *without a code*, in a self-organized, emergent fashion, like the autocatalytic sets of polymers widely believed to be the earliest form of life. These life forms generated self-similar structure, but since there was no code yet to copy from, there was no explicit copying going on. The presence of a given catalytic polymer, say polymer X, simply sped up the rate at which certain reactions took place, while another catalytic polymer, say Y, influenced the reaction that generated X. Eventually, for each polymer in the set, there existed a reaction that catalyzed it. Because the process occured in a piecemeal manner, through bottom-up interactions rather than a top-down genetic code, they replicated with low fidelity, and acquired characteristics were inherited. We can refer to this kind of structure as a *primitive replicator.*

A worldview has a similar structure and dynamics. Just as polymers catalyze reactions that generate other polymers, retrieval of an item from memory can trigger another, which triggers yet another and so forth, thereby cross-linking memories, ideas, and concepts into a conceptually-closed web. Thus a worldview constitutes a second kind of primitive replicator, and it is worldviews (not separate ideas or memes) that evolve. A worldview is not just a collection of discrete ideas or memes, nor do ideas or memes form an interlocking set like puzzle pieces, because each context impacts it differently, fragmenting it into a slightly different puzzle. So, in contradiction to the meme perspective, neither a painting nor the ideas that went through the artist's mind while painting it constitute a replicator. A painting plays its role in the evolution of culture by revealing some aspect of the artist's worldview (which *is* a replicator) and thereby affecting the worldviews (other replicators) of those who admire it.

As with the earliest forms of life, traits acquired over a lifetime are heritable; that is, get passed on from one 'generation' to the next. We hear a joke and, in telling it, give it our own slant, or we create a disco version of Beethoven's Fifth Symphony and a rap version of that. The evolutionary trajectory of a worldview makes itself known indirectly, like footprints in the sand, via the behavior and artifacts it manifests under the influence of the contexts it encounters. For example, when you explain how to change a tire, certain facets of your worldview are revealed, while playing a piano concerto reveals others. The situation of a flat tire 'sliced through' your worldview in such a way that certain parts of it were expressed, while the concerto expressed others.

Thus we argue that while *brains* were evolving through biological evolution, *conceptually closed worldviews* began evolving through cultural evolution. This second evolutionary process rides piggybacks on the first, and the two mutually reinforce one another. As worldviews become increasingly complex, the artifacts they manifest become increasingly complex, which necessitates even more complex worldviews, *et cetera*.

### *4.2 Creative Thought is Not a Darwinian Process*

Elsewhere we have presented forceful arguments that, contrary to some psychologists (Campbell, 1960, 1965, 1987; Simonton, 1999a, 1999b), neither worldviews nor the creative processes they generate evolve through natural selection (Gabora & Aerts, in press a). Selection theory requires multiple, distinct, simultaneously-actualized states. In cognition, each thought or cognitive state changes the 'selection pressure' against which the next is evaluated; they are not simultaneously selected amongst. Creative thought is more a matter of honing in on a vague idea by redescribing successive iterations of it from different real or imagined perspectives; in other words, actualizing potential through exposure to different contexts. It has been proven that the mathematical description of contextual change of state introduces a non-Kolmogorovian probability distribution, and a classical formalism such as selection theory cannot be used (Aerts, 1986; Accardi, 1982; Aerts & Aerts, 1994; Pitowsky, 1989; Randall & Foulis, 1976). Thus an idea certainly changes as it gets mulled over in a stream of thought, and indeed it appears to evolve, but the process by which it evolves is not Darwinian.

Natural selection as it has been mathematically formulated has been able to yield an approximate description of the evolution of biological organisms because self-replication instructions are encoded in the form of a genome, which is shielded from contextual influence; the genome of the child does not retain change acquired over the lifetime of the parent. However, this is not the case for cultural evolution and the cognitive processes underlying it (nor for the earliest forms of biological life itself). In a stream of thought, or a discussion amongst individuals, neither are all contexts equally likely, nor does context have a limited effect on future iterations. So the assumptions that make classical stochastic models useful approximations do not hold for creative thought. Attempts to apply selection theory to thought commit the serious error of treating a set of *potential*, contextually elicited states of *one* entity as if they were *actual* states of a *collection* of entities, or possible states with no effect of context, even though the mathematical structure of the two is completely different.

### *4.3 Evolution as Context-driven Actualization of Potential*

We have seen that human culture does appear to evolve, and examined two transitions in its evolution. However, we have also seen that the process through which it evolves is not strictly Darwinian. How then *does* this process of evolution work?

In fact, probing the similarities and differences between biological and cultural evolution can deepen our understanding of how *any* sort of evolutionary process could manifest itself. It is becoming increasingly evident that the Darwinian (or neo-Darwinian) paradigm, powerful though it is, does not even provide a comprehensive account of biological processes of change (*e.g*. Kauffman, 1993; Newman & Muller, 1999; Schwartz, 1999)[2] let alone nonbiological processes. There is no reason evolution must be Darwinian, or even involve selection except as a special case. It is not incorrect to use the term evolution in a more inclusive sense as adaptive change in response to environmental constraint; physicists use it to refer to change in the absence of a measurement, without implying that selection is involved. It may be that it is only because Darwinian evolution is such an *unusual* form of evolution that it got so much attention it eventually cornered the word 'evolution'.

We have been working on a general, transdisciplinary framework for the description and analysis of evolutionary processes (Gabora & Aerts, in press b). In a nutshell, evolution is viewed as process through which an entity actualizes its potential for change, sometimes through interaction with a context (*e.g*. stimulus, situation, or environment). In other words, it is a process of *context-driven actualization of potential*, or CAP. Different forms of evolution differ with respect to the degree to which they are sensitive to, internalize, and depend upon a particular context, and whether change of state is deterministic or nondeterministic.

The mathematical structure used to model the change of state of an entity through context-driven actualization of potential (whether it be quantum particle, macro object, or concept) is the State Context Property System (SCOP). A SCOP consists of a set of states $\Sigma$, a set of relevant contexts $M$, and a set of relevant properties $L$. A change of state modeled by SCOP is of the following form: an entity in a specific state $p$ in $\Sigma$ changes under the influence of a specific context $e$ in $M$ changes to another state $q$ in $\Sigma$. Each state has different applicabilities of properties and different probabilities of changing to each other state. It is this means of describing dynamic change under the influence of a context that allows the modeling of subject-object interaction. By way of enabling cross-disciplinary comparison, the CAP framework illustrates how unusual Darwinian evolution is, and clarifies in what sense culture is and is not Darwinian. Thus we reach a more general understanding of how it is that something could evolve.[3]

---

[2]For example, nonDarwinian processes such as self-organization, assortative mating, and epigenetic mechanisms play an important role in biology.

[3] The CAP framework also has implications for the 'hard' sciences. For example, it suggests that the dynamical evolution of a quantum entity is not fundamentally different from collapse, but rather a change of state for which there is only one way *to* collapse.

We will see shortly how this framework is used to model the change-of-state a mind undergoes as it evokes a concept in a particular context.

## 5 Concepts: An Enigma at the Heart of the Problem

We have seen that at the heart of the question of how culture evolves lies the question of how novelty is generated. And at the heart of *that* question lies the thorny problem of understanding the flexible way we use concepts. The rationale and philosophy underlying our approach to concepts is outlined in (Aerts *et al.,* 2004, in press; Gabora *et al*., 2005). Traditionally concepts were viewed as entities in the mind that *represent* a class of entities in the world. However, it has been pointed out that they do not have a fixed representational structure; the relevance or applicability features or properties changes depending on the context in which the concept arises (Rosch, 1973; Barsalou, 1982). (For example, although the concept *baby* can refer to a real human baby, a plastic doll, or a stick figure painted with icing on a cake, for each situation the set of properties is different.) Indeed when concepts combine, certain properties disappear altogether, while new ones come into play (*e.g*. when *baby* combines with *doll* in the conjunction *baby doll*, properties atypical of *baby*, such as 'made of plastic', are gained, while other *baby* properties such as 'has DNA', are lost). So although until recently the primary function of concepts was thought to be the *identification* of items as instances of a particular class, increasingly they are thought to not just identify but *actively participate* in the generation of meaning (Rosch 1999). Thus a complete theory of concepts requires a mathematical formalism that can describe the contextuality with which they adapt to situations. It must transcend the Cartesian worldview in which an entity is viewed as separate and distinct from the environment it inhabits.

Our approach follows naturally from previous research on the generalization of the mathematical formalisms of quantum mechanics, and application of these generalized formalisms to other situations involving contextuality, allowing us to incorporate the effect of a measurement or context into the description of the entity itself. As outlined in (Gabora & Aerts 2002), the situation concepts research faces now is reminiscent of the situation encountered in physics a century ago. Quantum mechanics was born when experiments on micro-particles revealed, for the first time in history, a world that resisted description using the mathematics of classical mechanics which had until then been completely successful. Like quantum entities whose manifestations change depending on the measurement context, a concept's manifestation also depends on the situation (context) in which it is encountered. Prior to the measurement or context, both quantum entities and concepts can be described as existing in a state of *potentiality* with respect to this measurement or context, which mathematically — if the state space is the linear vectorspace of quantum mechanics — is a *superposition* of the different states it can change to under influence of the measurement or context. As in quantum mechanics, where the applicability of a property depends on the context of a measurement relevant to the detection of that property, the applicabilities of features of a concept also depend on the particular context in which it is evoked. As in quantum mechanics, where it is common for two entities to combine to become one, it is common for concepts to spontaneously combine to form conjunctions of concepts. Quantum mechanics provides a means of mathematically describing the process whereby two entities merge to become one, and generalizations of these formalisms are applicable to not just entangled quantum particles but also conjunctions of concepts. (Note that this kind of re-application of a generalized mathematical structure has nothing to do with investigations of how phenomena at the quantum level affect cognition.)

### *5.1 The SCOP Representation of a Concept*

Consider two contexts for the concept 'pet': 'The pet is chewing a bone' and 'The pet is being taught to talk'. In (Aerts & Gabora 2005a) we show that if subjects are asked to rate the typicality of a specific exemplar of 'pet' (e.g. *dog*) and the applicability of a particular property of 'pet' (e.g. *furry*), their ratings will depend on whether 'pet' is considered under the first context or the second. Thus *dog* rates high under the first context and low under the second, whereas *parrot* shows the inverse pattern.

Similarly with properties, *furry* rates high under the first context and low under the second, whereas *feathered* shows the inverse pattern. A basic aim of our formalism is to model this type of contextual influence. We begin by introducing the notion of 'state of a concept'. When a concept is not being considered under any particular context, and indeed not the subject of conscious thought, we refer to it as being in its *ground state* or unexcited state. When a concept is evoked by some context, we refer to the concept as being in an *excited state*. Each context manifests a different excited state, and each excited state is associated with different exemplar typicalities and property applicabilities. Note that we are not just proposing that the applicabilities of properties differ for different exemplars of a concept, an effect accounted for in other theories, *e.g*. prototype and exemplar theories. The applicability of a single property varies for each state, as does the typicality of a single exemplar. Thus for the above example we introduce two states of the concept 'pet', *i.e*. one that accounts for the ratings under the first context, and another that accounts for the ratings under the second.

Other theories of concepts have difficulty accounting for why items that are dissimilar or even opposite might nevertheless belong together; for example, why *white* might be more likely to be categorized with *black* than with *flat*, or why *dwarf* might be more likely to be categorized with *giant* than with, say, *salesman*. Adopting the quantum terminology, this problem gets solved by distinguishing between similarity with respect to which contexts are relevant—*compatibility*—and similarity with respect to values for those contexts—*correlation*. This refined notion of similarity enables us to develop context-sensitive measures of conceptual distance.

We develop the mathematical structure of a concept using SCOP by identifying structures of the sets of states $\Sigma$, contexts $M$, and properties $L$. Each state $p$ in $\Sigma$ has its own typicality values for exemplars and applicabilities of properties. Consider for the concept 'pet', contexts $e$ 'The pet runs quickly through the garden' and $f$ 'the pet runs quickly', we can say that $e$ 'is stronger than or equal to' $f$, thereby introducing a partial order relation in the set of contexts $M$. By introducing the 'and' context and the 'or' context, set $M$ obtains the structure of a complete lattice. By introducing the 'not' context for any other context, the structure of an orthocomplementation can be derived for $M$. If the state of a concept is not affected by a context it is said to be an *eigenstate* for this context. Otherwise it is a *potentiality state* for this context (reflecting its susceptibility to change).

We find that the structure of the SCOP for a concept entails a nonclassical (i.e. quantum) logical structure. One manifestation of this is that for two contexts $e$ and $f$ in $M$, a state is not necessarily an eigenstate of the context $e$ 'or' $f$ if and only if it is an eigenstate of $e$ 'or' an eigenstate of $f$. Another such manifestation is that although any state is an eigenstate of the context $e$ 'or' *not* $e$, we cannot say that any state is an eigenstate of $e$ 'or' an eigenstate of *not* $e$. The latter argument can easily be illustrated by the contrast between context $e$ 'The pet runs quickly', and context *not* $e$ 'It is not so that the pet runs quickly'. Any state of 'pet' that does not specifically refer to what the pet is doing is neither an eigenstate of $e$ (indeed, in $e$ the state of pet changes to one in which the pet 'runs quickly') nor is it an eigenstate of *not* $e$ (here the state of 'pet' is also affected; it changes to a state in which the pet 'it is not so that the pet runs quickly'). A complete orthocomplemented lattice structure is also derived for the set of properties $L$. The existence of a complete lattice structure for the sets of contexts and properties makes it possible to construct a topological representation of a SCOP in a closure space.

### *5.2 Embedding the SCOP in Complex Hilbert Space*

The identification of the complete orthocomplemented lattice structure for the sets of contexts and properties of the SCOP is an operational derivation, *i.e*. we do not make any non-operational technical hypothesis, but merely derive the structure by taking into account the natural relations (such as the partial order relation of 'stronger than or equal to') that exist in the sets of contexts and properties. We have also taken a non-operational step, embedding the SCOP in a more constrained structure, the complex Hilbert space, the mathematical space used in quantum mechanical formalism. We have good reason to do so. The generalized quantum formalisms entail the structure of a complete orthocomplemented lattice, and its concrete form, standard quantum mechanics, is formulated within a complex Hilbert space. By formulating the SCOP representation of a concept in terms of the much less abstract numerical space, the complex Hilbert space, we make strong gains in terms of calculation and prediction power. For the mathematics of a standard quantum mechanical model, it is not only the vector space structure of the complex Hilbert space that is important, but also the way the Hilbert space is used. A state is described by a unit vector or a density operator, and a context or property by an orthogonal projection. The quantum formalism furthermore determines the formulas that describe the transition probabilities between states and the weights of properties. These allow us to model the typicality of exemplars and applicability of properties. Predictions of frequency values of exemplars and applicability values of properties coincided with the values yielded by the experiment for the concept 'pet'. Thus theoretical and experimental results indicate that our approach successfully describes the contextual manner in which concepts are used.

### *5.3 Concept Combination*

Probably the most drastic illustration of how concepts shift in meaning in different contexts is when they combine to form a *conjunction*. It is well known that the typicality of the conjunction is not a simple function of the typicality of its constituents. This has come to be called the 'pet fish problem' because of the well-known example where *guppy* is rated as a good example, not of the concept 'pet', nor of the concept 'fish', but of the conjunction 'pet fish'. This phenomenon has resisted explanation by contemporary theories of concepts because they have no means to describe dynamical change of state under the influence of a context, thus they cannot describe the change of state that concepts evoke in one another when they act as contexts for each other. Because this kind of dynamical change of state is a fundamental component of the SCOP formalism, this problem disappears.

In (Aerts & Gabora, 2005b) we take 'The pet is a fish' to be a context for the concept 'pet', and 'The fish is a pet' to be a context for the concept 'fish', and look at the change of state they mutually provoke in one another. The mathematical structure used to describe a compound of two quantum entities is the tensor product of each of their individual Hilbert spaces. It generates elements called *non-product vectors*, which describe *states of entanglement* that spontaneously come into existence when entities combine, and which can exhibit a gain or loss of properties or quite different properties altogether from the states of the constituent entities. It is these non-product states that describe conjunctions (*e.g*. 'pet fish' is described as an entangled states of the concepts 'pet' and 'fish'). The tensor product procedure also allows the modeling of more complex combinations of concepts such as 'a pet and a fish' (which is completely different from 'pet fish'). In this case, product states are involved, which means that the combining of concepts employing the word 'and' does not entail entanglement. Using as an example the sentence 'The cat eats the food', we have shown how our theory makes it possible to describe the combination of an arbitrary number of concepts.

## 6 Summary and Conclusions

If we are to take seriously the idea that culture is an evolutionary process, we can indeed look to evolution to provide the kind of overarching framework for the humanities that it provides for the biological sciences. But in so doing, we must come to a more general view of what evolution can be and what it involves—context-driven actualization of potential. Although *some* aspects of culture are amenable to Darwinian description, such as the competition of artifacts in the marketplace, others, such as the origin of culture, require concepts from complexity theory such as self-organization and emergence. Still other aspects of culture, particularly those that involve the generation and refinement of novel ideas within and between individuals, require for their formal description a means of dealing with potentiality, context, and nondeterminism.

We saw that it is possible to evolve stuff using a cultural algorithm that simulates invention and imitation. But what is most elusive about culture is how ideas change by examining them from different perspectives, alone or in a group, and how change to one idea percolates to other ideas due to their mutual influence on one another as different facets of an integrated worldview. We have argued that the origin of culture can be attributed to achievement of such an integrated worldview and explain how this could have come about through a process of conceptual closure. We suggest that the cultural transition of the Middle-Upper Paleolithic, hailed as period that gave birth to art, science, and religion, resulted from onset of contextual focus: the capacity to spontaneously focus or defocus attention, thereby alternating between analytical and associative thought, and enhancing effective traversal and integration of the worldview. Moreover, we argue that an integrated worldview constitutes a primitive replicator, similar to the self-organized autocatalytic sets postulated to be the earliest forms of life. Like the first living things, it is self-organized and self-mending, and its replication is emergent rather than dictated by self-assembly instructions (such as the genetic code) and therefore subject to inheritance of acquired characteristics (*e.g*. if one modifies a joke, friends may well pass it on to others in modified form). Thus it is not ideas, artifacts, or 'memes' that evolve, but minds. Different contexts expose different facets of a mind (like cutting a fruit at different angles exposes different parts of its interior). If you are friendly or Christian or aesthetically inclined, that will manifest in many different ways under different circumstances as words, actions, gestures, and facial expressions. If we are right, we it is not these manifestations that are evolving, but the worldviews that underlie them. The words and actions are merely the mind's way of revealing its current evolutionary state.

The key to understanding the generativity, adaptability, and creativity of human thought is to understand how concepts are represented in the mind such that they can be flexibly evoked and applied to different situations. Science has had difficulty accounting for the fluidity and compositionality of concepts in everyday situations. We have proposed a theory for modeling concepts that uses the state-context-property theory (SCOP), a generalization of the quantum formalism, whose basic notions are states, contexts and properties basic notions are states, contexts and properties. A concept (*e.g*. 'cup') is defined not just in terms of exemplary instances or states (*e.g*. 'tea cup') and their features or properties (*e.g*. 'concave'), but also by the relational structures of these properties, and their susceptibility to change under different contexts. We differentiate between *potentiality states with respect to a context* of a concept, which have the potential to undergo a change of state with respect to this context, and *eigenstates with respect to a context*, which do not. SCOP enables us to incorporate context into the mathematical structure used to describe a concept, and thereby model how context influences the typicality of an instance or exemplar and the applicability of a property. It is also possible to embed the sets of contexts and properties of a concept in the complex Hilbert space of quantum mechanics. This enables conjunctions to be described as states of entanglement using the tensor product. The mathematics extends readily to more complex or combinations of concepts, with clear implications for the productivity of spoken and written language.

Let us end by returning to the everyday questions that can inspire this kind of investigation. Specifically, one can, say, buy something nice and feel happy for a while, but sooner or later life feels hollow unless there is a sense of purpose, a sense of one's place in something larger, an effect one can have. By viewing each person as a conscious player in the process through which worldviews evolve, each act becomes sacred and imbued with the potential to change the future. This is what it really means to be human, to partake in the web of thoughts and cultural artifacts that extends back to our earliest ancestors and will affect all who come next. Maladaptive worldviews have a chance to get replaced by (literally) more evolved ones, and there are mechanisms at work to help this along. We feel happy when we recognize ideas or attitudes that are clearly emanations of a beautiful worldview. Such experiences guide us as individuals to take meaningful steps toward a second form of evolution, cultural evolution, which got started much later than biological evolution, but is every bit as remarkable. It is the process through which today's culture is rooted in cultures of the past, the process whereby our thoughts generate actions, which touch others, which touch still others, and thus a vast web of conscious minds together weave the fabric of their reality, forever creating new ways of seeing and being.

## ACKNOWLEDGEMENTS

This research was supported by Grant G.0339.02 of the Flemish Fund for Scientific Research. We also thank the *Foundation for the Future* for making this most interesting workshop possible.

## REFERENCES

Accardi L. & Fedullo A. (1982) On the statistical meaning of complex numbers in quantum mechanics, *Il Nuovo Cimento*, **34**(7): 161-172.

Aerts, D. (1986) A possible explanation for the probabilities of quantum mechanics, *Journal of Mathematical Physics*, **27**: 202-210.

Aerts, D., & Aerts, S. (1994) Application of quantum statistics in psychological studies of decision processes, *Foundations of Science*, **1**: 85-97.

Aerts, D. & Gabora, L. (2005) A state-context-property model of concepts and their combinations I: The structure of the sets of contexts and properties. *Kybernetes,* **34**(1&2), special issue dedicated to Heinz Von Foerster.

Aerts, D. & Gabora, L. (2005) A state-context-property model of concepts and their combinations II: A Hilbert space representation. *Kybernetes,* **34**(1&2), special issue dedicated to Heinz Von Foerster.

Ashby, F. G., & Ell, S. W. (2002) Single versus multiple systems of learning and memory. In J. Wixted & H. Pashler (Eds.) *Stevens' handbook of experimental psychology: Volume 4 Methodology in experimental psychology*. New York: Wiley.

Aunger R. (2000) *Darwinizing Culture: The Status of Memetics as a Science*, Oxford University Press, Oxford.

Baldassarre, G. (2001) Cultural Evolution of 'Guiding Criteria' and Behaviour in a Population of Neural-network Agents, *Journal of Memetics,* **4**.

Barron, F. (1963) *Creativity and Psychological Health*, Van Nostrand.

Barsalou, L. W. (1982) Context-independent and context-dependent information in concepts, *Memory & Cognition*, **10**: 82-93.

Blackmore, S. (1999) *The Meme Machine*, Oxford University Press, Oxford.

Blackmore, S. (2000) The Power of Memes, *Scientific American,* **283**(4): 52-61.

Boyd, R. & Richerson, P. J. (1985) Culture and the evolutionary process. University of Chicago Press.

Campbell D. T. (1965) Variation and selective retention in socio-cultural evolution. In H. R. Barringer, G. I. Blanksten, & R. W. Mack (Eds.), *Social Change in Developing Areas: A Reinterpretation of Evolutionary Theory*. Cambridge, MA Schenkman.

Campbell, D. T. (1987) Evolutionary epistomology. In G. Radnitzky & W. W. Bartley III (Eds.), *Evolutionary Epistomology, Rationality, and the Sociology of Knowledge*. LaSalle, IL: Open Court.

Campbell, D.T. (1990) Levels of organization, downward causation, and the selection-theory approach to evolutionary epistemology. In Greenberg & Tobach (Eds.), *Theories of the Evolution of Knowing*. Hillsdale NJ: Erlbaum.

Cavalli-Sforza, L. L. & Feldman, M. W. (1981) *Cultural Transmission and Evolution: A Quantitative Approach*, Princeton University Press.

Churchland, P. S. and Sejnowski, T. (1992) *The computational brain.* Cambridge MA: MIT Press.

Davis, J., & Rovee-Collier, C. (1983) Alleviated forgetting of a learned contingency in 8-week-old infants. *Developmental Psychology,* **19**: 353-365.

Dawkins, R. (1976) *The Selfish Gene*, Oxford University Press, Oxford.

Edelman, G. (2000) Bright air, brilliant fire: On the matter of the mind. New York: Basic Books.

Erdos, P. & Renyi, A. (1959) On the random graphs 1(6), Institute of Mathematics, University of Debrecenians, Debrecar, Hungary.

Erdos, P. & Renyi, A. (1960) On the evolution of random graphs. *Institute of Mathematics, Hungarian Academy of Sciences Publication* number 5.

Fauconnier, G. & Turner, M. (2002) *The way we think: Conceptual blending and the mind's hidden complexities.* New York: Basic Books.

Freud, S. (1949) *An outline of psychoanalysis*. New York: Norton.

Eysenck, H. J. (1995) *Genius: The Natural History of Creativity*, Cambridge University Press, Cambridge UK.

Feist, G. J. (1999) The influence of personality on artistic and scientific creativity. In: *Handbook of Creativity,* ed. R. J. Sternberg, Cambridge University Press, Cambridge UK, 273-296.

Finke, R. A., Ward, T. B. & Smith, S. M. (1992) *Creative cognition: Theory, research and applications*. Cambridge MA: MIT Press.

Fodor, E. M. (1995) Subclinical manifestations of psychosis-proneness, ego-strength, and creativity. *Personality and Individual Differences* **18**: 635-642.

Gabora, L. (1995) Meme and variations: A computer model of cultural evolution. In (L. Nadel & D. Stein, Eds.) 1993 Lectures in Complex Systems, Addison-Wesley, 471-486.

Gabora, L. (1996) A day in the life of a meme. *Philosophica* **57**: 901-938.

Gabora, L. (1997) The origin and evolution of culture and creativity. *Journal of Memetics: Evolutionary Models of Information Transmission* **1**(1).

Gabora, L. (1998) Autocatalytic closure in a cognitive system: A tentative scenario for the origin of culture. Psycoloquy **9**(67). http://www.cogsci.ecs.soton.ac.uk/cgi/psyc/newpsy?9.67

Gabora, L. (1999) Weaving, bending, patching, mending the fabric of reality: A cognitive science perspective on worldview inconsistency. *Foundations of Science* **3**(2): 395-428.

Gabora, L. (2000) Toward a theory of creative inklings. In (R. Ascott, Ed.) *Art, Technology, and Consciousness*, Intellect Press, 159-164.

Gabora, L. (2000) Conceptual closure: Weaving memories into an interconnected worldview. In (G. Van de Vijver & J. Chandler, Eds.) *Closure: Emergent Organizations and their Dynamics*. Vol. 901 of the Annals of the New York Academy of Sciences, 42-53.

Gabora, L. (2002) The beer can theory of creativity. In (P. Bentley & D. Corne, Eds.) *Creative Evolutionary Systems*, Morgan Kauffman, 147-161.

Gabora, L. (2002) Cognitive mechanisms underlying the creative process. In (T. Hewett and T. Kavanagh, Eds.) *Proceedings of the Fourth International Conference on Creativity and Cognition*, 13-16 October, Loughborough University UK, 126-133.

Gabora, L. (2002) Amplifying phenomenal information: Toward a fundamental theory of consciousness. *Journal of Consciousness Studies* **9**(8): 3-29.

Gabora, L. (2003) Contextual focus: A cognitive explanation for the cultural transition of the Middle/Upper Paleolithic. *Proceedings of the 25th Annual Meeting of the Cognitive Science Society,* Boston MA, 31 July-2 August. Lawrence Erlbaum Associates.

Gabora, L. (2004) Ideas are not replicators but minds are. *Biology and Philosophy* **19**(1): 127-143.

Gabora, L. & Aerts, D. (2002) Contextualizing concepts using a mathematical generalization of the quantum formalism. *Journal of Experimental and Theoretical Artificial Intelligence* **14**(4): 327-358.

Gabora, L. & Aerts, D. (in press) Creative thought as a non-Darwinian evolutionary process. *Journal of Creative Behavior*.

Gabora, L. & Aerts, D. (in press) Evolution as context-driven actualization of potential. *Interdisciplinary Science Reviews*.

Gabora, L., Rosch, E. & Aerts, D. (2005) Toward an ecological theory of concepts. In (D. Aerts, B. D'Hooghe & N. Note, Eds.) *Worldviews, Science and Us: Bridging Knowledge and Perspectives on the World*, World Scientific, Singapore.

Holland, J. (1975) *Adaptation in Natural and Artificial Systems*. University of Michigan Press, Ann Arbor.

Howard-Jones, P. A. & Murray, S. (2003) Ideational productivity, focus of attention, and context. *Creativity Research Journal* **15**(2&3): 153-166.
James, W. (1890/1950). *The principles of psychology.* New York: Dover.
Johnson-Laird, P.N. (1983). *Mental models*. Cambridge MA: Harvard University Press.
Karmiloff-Smith, A. (1992) *Beyond modularity: A developmental perspective on cognitive science*, Cambridge MA: MIT Press.
Kauffman, S. A. (1993) *Origins of order*. New York: Oxford University Press.
Kauffman, S. A. (1999) Darwinism, neoDarwinism, and the autocatalytic model of culture: Commentary on Origin of Culture, *Psycoloquy* **10**(22): 1-4.
Kris, E. (1952) *Psychoanalytic explorations in art*. New York: International Universities Press.
Martindale, C. (1995) Creativity and connectionism. In S. M. Smith & T. B. Ward, & R. A. Finke (Eds.) *The creative cognition approach* (pp. 249-268). Cambridge MA: MIT Press.
Matzel, L., Collin, C., & Alkon, D. (1992) Biophysical and behavioral correlates of memory storage: Degradation and reactivation. *Behavioral Neuroscience,* **106**: 954-963.
Neisser, U. (1963). The multiplicity of thought. *British Journal of Psychology*, 54, 1-14.
Pitowsky, I. (1989) *Quantum Probability - Quantum Logic: Lecture Notes in Physics 321*, Berlin: Springer.
Randall, C. & Foulis, D. (1976). A mathematical setting for inductive reasoning. In C. Hooker (Ed.), *Foundations of Probability Theory, Statistical Inference, and Statistical Theories of Science* (vol. III, pp. 169). Dordrecht: Kluwer Academic.
Richards, R.L., Kinney, D.K., Lunde, I., Benet, M., & Merzel, A. (1988) Creativity in manic depressives, cyclothymes, their normal relatives, and control subjects. *Journal of Abnormal Psychology* **97**: 281-289.
Rips, L. (2001) Necessity and natural categories. *Psychological Bulletin,* **127**(6): 827-852.
Rosch, E. (1999) Reclaiming concepts, *Journal of Consciousness Studies*, **6**: 61-78.
Rovee-Collier, C., Hartshorn, K., & DiRubbo, M. (1999) Long-term maintenance of infant memory. *Developmental Psychobiology,* **35**: 91-102.
Russ, S. W. (1993) *Affect and Creativity*. Erlbaum, Hillsdale NJ.
Simonton, D. K. (1999a) *Origins of genius: Darwinian perspectives on creativity*. New York: Oxford.
Simonton, D. K. (1999b) Creativity as blind variation and selective retention: Is the creative process Darwinian? *Psychological Inquiry* **10**: 309-328.
Sloman, S. (1996) The empirical case for two systems of Reasoning. *Psychological Bulletin,* **9**(1): 3-22.
Smolensky, P. (1988) On the proper treatment of connectionism. *Behavioral and Brain Sciences,* **11**: 1-43.
Spector, L. & Luke, S. (1996a) Culture Enhances the Evolvability of Cognition, *Proceedings of the Eighteenth Annual Conference of the Cognitive Science Society*, G. Cottrell (ed.), Lawrence Erlbaum Associates, Mahwah, NJ, 672-677.
Spector, L. & Luke, S. (1996b) Cultural Transmission of Information in Genetic Programming, in *Genetic Programming 1996: Proceedings of the First Annual Conference*, J. R. Koza, D. E. Goldberg, D. B. Fogel, and R. L. Riolo (eds.), MIT Press, Cambridge MA, 209-214.
Tomasello, M., A. Kruger, A. & Ratner, H. (1993) Cultural learning. *Behavioral and Brain Sciences* **16**: 450-88.
Von Neumann, J. (1966) *Theory of Self-reproducing Automata,* University of Illinois Press, Champaign.